\documentclass[10pt,letterpaper]{article}
\usepackage[top=0.85in,left=2.75in,footskip=0.75in]{geometry}

\usepackage{amsmath,amssymb}

\usepackage{changepage}

\usepackage{textcomp,marvosym}

\usepackage{cite}

\usepackage{nameref,hyperref}

\usepackage[nopatch=eqnum]{microtype}
\DisableLigatures[f]{encoding = *, family = * }

\usepackage[table]{xcolor}

\usepackage{array}

\newcolumntype{+}{!{\vrule width 2pt}}

\usepackage{tikz}
\usepackage{tikz-cd}
\usetikzlibrary{trees}
\usepackage{verbatim}
\usepackage{quoting} 
\usepackage{units}
\usepackage{xurl}
\usepackage{pgfplots}
\pgfplotsset{compat=newest}
\pgfplotsset{compat=1.17}
\usepackage{filecontents}
\usepackage{xcolor}
\newlength\savedwidth

\newtheorem{definition}{Definition}

\raggedright
\usepackage[aboveskip=1pt,labelfont=bf,labelsep=period,justification=raggedright,singlelinecheck=off]{caption}

\makeatletter
\renewcommand{\@biblabel}[1]{\quad#1.}
\makeatother

\usepackage{lastpage,fancyhdr,graphicx}
\usepackage{subcaption}
\usepackage{epstopdf}
\fancyheadoffset[L]{2.25in}
\fancyfootoffset[L]{2.25in}
\begin{document}
\vspace*{0.2in}

\begin{flushleft}
{\Large
\textbf\newline{AI-powered mechanisms as judges: Breaking ties in chess\footnote{Accepted at PLOS ONE.}} 
}
\newline

Nejat Anbarci\textsuperscript{1\Yinyang},
Mehmet S Ismail\textsuperscript{2*\Yinyang}
\\
\bigskip
\textbf{1} Department of Economics, Durham University, Durham DH1 3LB, UK
\\
\textbf{2} Department of Political Economy, King's College London, London, WC2R 2LS, UK
\\
\bigskip

%
%
\Yinyang These authors contributed equally to this work.




* mehmet.ismail@kcl.ac.uk

\end{flushleft}
\section*{Abstract}
Recently, Artificial Intelligence (AI) technology use has been rising in sports to reach decisions of various complexity. At a relatively low complexity level, for example, major tennis tournaments replaced human line judges with Hawk-Eye Live technology to reduce staff during the COVID-19 pandemic. AI is now ready to move beyond such mundane tasks, however. A case in point and a perfect application ground is chess. To reduce the growing incidence of ties, many elite tournaments have resorted to fast chess tiebreakers. However, these tiebreakers significantly reduce the quality of games. To address this issue, we propose a novel AI-driven method for an objective tiebreaking mechanism. This method evaluates the quality of players' moves by comparing them to the optimal moves suggested by powerful chess engines. If there is a tie, the player with the higher quality measure wins the tiebreak. This approach not only enhances the fairness and integrity of the competition but also maintains the game's high standards. To show the effectiveness of our method, we apply it to a dataset comprising approximately 25,000 grandmaster moves from World Chess Championship matches spanning from 1910 to 2018, using Stockfish 16, a leading chess AI, for analysis.


\section{Introduction}\label{sec1}

The use of Artificial Intelligence (AI) technology in sports, to reach decisions of various complexity, has been on the rise recently. At a relatively low complexity level, for example, major tennis tournaments replaced human line judges with Hawk-Eye Live technology to reduce staff during the COVID-19 pandemic. Also, more than a decade ago, football began using Goal-line technology to assess when the ball has completely crossed the goal line. These are examples of mechanical AI systems requiring the assistance of electronic devices to determine the precise location of balls impartially and fairly, thus minimizing, if not eliminating, any controversy.

\begin{figure}
    \centering
    \includegraphics[width=0.65\textwidth]{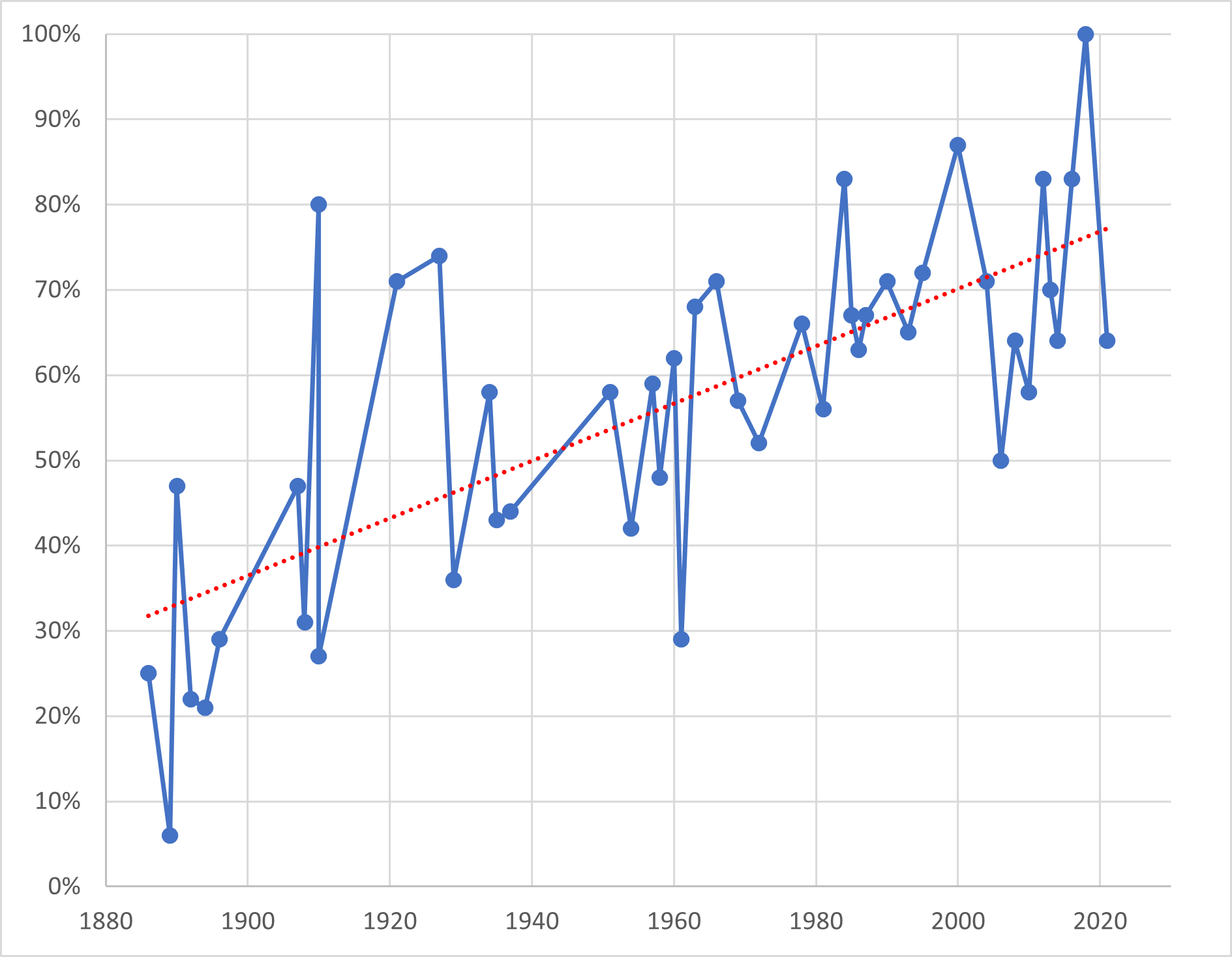}
    \vspace{0.1cm}
    
    \caption{Percentage of draws in World Chess Championship matches 1886--2021.}
    \label{fig:wch}
\end{figure}

A major question now is whether AI could move beyond such rudimentary tasks in sports. A case in point and a perfect application ground is chess for two complementary reasons. On the one hand, advanced AI systems, including Stockfish, AlphaZero, and MuZero, have already been implemented in chess \cite{stockfish,silver2018,schrittwieser2020}; further, the superiority of top chess engines has been widely acknowledged ever since IBM's Deep Blue defeated former world chess champion Garry Kasparov in 1997 \cite{campbell2002}. On the other hand, despite its current popularity all around the world, chess is very much hampered by the growing incidence of draws, especially in the world championships, as Fig \ref{fig:wch} illustrates. In response to this draw problem, elite chess tournaments---like other sports competitions \cite{anbarci2021,apesteguia2010,berker2014,brams2018,brams2021,cohen2018,csato2023}---resorted to the tiebreakers. The most common final tiebreaker is the so-called Armageddon game, where White has more time (e.g., five minutes) to think on the clock than Black (e.g., four minutes), but Black wins in the event of a draw. However, this format sparks controversy among elite players and chess aficionados alike:
\begin{quote}
``Armageddon is a chess penalty shoot-out, a controversial format intended to prevent draws and to stimulate interesting play. It can also lead to chaotic scrambles where pieces fall off the board, players bang down their moves and hammer the clocks, and fractions of a second decide the result'' (Leonard Barden, \textit{The Guardian} \cite{barden2019}).
\end{quote}

\begin{table}[h!]
\centering
\[
\arraycolsep=1.1pt\def\arraystretch{1.5}
\begin{array}{ r|c|c|}
\multicolumn{1}{r}{}
&  \multicolumn{1}{c}{\text{Time-control}}
& \multicolumn{1}{c}{\text{Time per player}}\\
\cline{2-3}
&  \text{Classical}  & \text{90 min. + 30 sec. per move}  \\
\cline{2-3}
&  \text{Rapid}  & \text{15 min. + 10 sec. per move}  \\
\cline{2-3}
&  \text{Blitz}  & \text{5 min. + 2 sec. per move} \\
\cline{2-3}
& ~ \text{Armageddon} ~ & ~\text{White: 5 min., Black: 4 min. (and draw odds)}  ~ \\
\cline{2-3}
\end{array}
\]
\caption{An example of classical vs. fast chess time-controls}
\label{table:time-control}
\end{table}

In this paper, we introduce an AI-based tiebreaking method designed to address the prevalence of draws in chess, particularly evident in World Chess Championship matches. This novel and practical approach, applied to a dataset of about 25,000 moves from grandmasters in matches spanning from 1910 to 2018, employs Stockfish 16, a leading chess AI system, to analyze each move. By assessing the difference between a player's actual move and the engine's optimal move, our method identifies the player with superior play quality in tied matches, thereby effectively resolving all such ties.

The ``draw problem'' that we address in this paper has a long history. Neither chess aficionados nor elite players appear to enjoy the increasing number of draws in chess tournaments. The five-time world champion, Magnus Carlsen, appears to be dissatisfied as well. ``Personally, I'm hoping that this time there will be fewer draws than there have been in the last few times, because basically I have not led a world championship match in classical chess since 2014'' \cite{carlsen2021}.

The 2018 world championship tournament, for instance, ended with 12 consecutive draws. The world champion was then determined by a series of ``rapid'' games, whereby players compete under significantly shorter time-control than the classical games (see Table~\ref{table:time-control}). If the games in the tiebreaks did not determine the winner, then a final game called Armageddon would have been played to determine the winner. Clearly, compared to classical chess games, there is no doubt that the fast-paced rapid, blitz and Armageddon games lower the quality of chess played; the latter also raises questions about its fairness because it treats players asymmetrically. (For fairness considerations in sports, apart from the works we have already mentioned, see, e.g., \cite{arlegi2020,csato2021}.)

More broadly, our paper relates to the literature on fairness of sports rules. Similar to the first-mover advantage in chess, there is generally a first-kicker advantage in soccer penalty shootouts \cite{apesteguia2010,rudi2020}. Several authors have proposed rule changes to restore (ex-post) fairness in penalty shootouts \cite{palacios2012,brams2018,anbarci2021,csato2022fairness,brams2023}.

The rest of the paper is organized as follows. In Section~\ref{sec:summary}, we give a brief summary of the empirical results. In Section~\ref{sec:chess}, we provide the main framework of the paper, discuss data collection, and describe the computation method. The limitations of our paper are discussed in Section~\ref{sec:limitations}; Section~\ref{sec:conclusions} concludes the paper.

\section{The empirical summary and the computation method}
\label{sec:summary}

In this paper, as mentioned, we introduce a novel and practicable AI-based tiebreaking method and show that it essentially eliminates the tie problem in chess. We apply this method to a dataset comprising approximately 25,000 moves made by grandmasters in World Chess Championship matches from 1910 to 2018. Notably, nine (about 18\%) out of fifty championship matches ended in a tie, with the incidence of ties increasing to 42\% in the past two decades.

In a nutshell, our analysis involved examining each move using Stockfish 16, one of the most advanced chess AI systems. We measured the discrepancy between the evaluations of a player's actual move and the optimal move as determined by the chess engine. In situations where matches ended in a tie, the player demonstrating a higher quality of play, as indicated by our analysis, is declared the winner of the tiebreak.

\subsection{An AI-based tiebreaking mechanism}

\begin{table}\[
\arraycolsep=1.1pt
\def\arraystretch{1.4}
\begin{array}{|c|c|c|c|c|c|c|}
\hline
~\text{Year}~ & ~\text{Players}~ & ~\text{\# Games}~ & ~\text{C-TPLV 1}~ & ~\text{C-TPLV 2} ~&~~~~ \Delta ~~~~ & \text{ ~~~\%~~~} \\
\hline
1910 & \text{Schlechter-Lasker} & 10 & 51.3 & 151.5 & -100.2 & 195.3 \\
\hline
1951 & \text{Botvinnik-Bronstein} & 24 & 245.4 & 145.1 & 100.3 & 69.1 \\
\hline
1954 & \text{Botvinnik-Smyslov} & 24 & 115.2 & 119.1 & -3.9 & 3.4 \\
\hline
1987 & \text{Kasparov-Karpov} & 24 & 93.9 & 88.8 & 5.1 & 5.7 \\
\hline
2004 & \text{Leko-Kramnik} & 14 & 117.1 & 22.4 & 94.7 & 422.8 \\
\hline
2006 & \text{Topalov-Kramnik} & 12 & 41.2  & 229.8& -188.6 & 457.8 \\
\hline
2012 & \text{Anand-Gelfand} & 12 & 42.3 & 43.9 & -1.6 & 3.8 \\
\hline
2016 & \text{Karjakin-Carlsen} & 12 & 41.9 & 40.6 & 1.3 & 3.2 \\
\hline
2018 & \text{Carlsen-Caruana} & 12 & 27.5 & 27.8 & -0.3 & 1.1 \\
\hline
\end{array}
\]
\caption{Summary statistics: The player with the lower cumulative (C) TPLV is declared the tiebreak winner. For details, see Fig~\ref{fig:all}. $\Delta$ represents the numerical difference and \% represents the percentage difference (with respect to the lower value) between the two C-TPLVs.}
\label{table:summary}
\end{table}

In the event of a win, it is straightforward to deduce that the winner played a higher quality chess than the loser. In the event of a tie in a tournament, however, it is more difficult to assert that the two players' performances were comparable, let alone identical. With the advancements in chess AIs, their differences in quality can be quantified. Average centipawn loss is a known metric for evaluating the quality of a player's moves in a game where the unit of measurement is 1/100th of a pawn. (See, e.g., \url{https://www.chessprogramming.org/Centipawns}.)

We instead use a more intuitive modification of that metric, which we term the ``total pawn loss,'' because (i) even chess enthusiasts do not seem to find the average centipawn loss straightforward, based on our own anecdotal observations, and (ii) it can be manipulated. The second point is easy to confirm because a player can intentionally extend the game, for example, in a theoretically drawn position, thereby `artificially' decreasing his or her average centipawn loss.

We define total pawn loss as follows. First, at each position in the game, the difference between the evaluations of a player's actual move and the ``best move'' as deemed by a chess engine is calculated. Then, the total pawn loss value (TPLV) for each player is simply the equivalent of the total number of ``pawn-units'' the player has lost during a chess game as a result of errors. TPLV can be calculated from the average centipawn loss and the number of moves made in a game (see Section~\ref{sec:chess}). If the TPLV is equal to zero, then every move was perfect according to the chess engine used. The lower the TPLV the better the quality of play. 

Table~\ref{table:summary} shows the cumulative TPLVs in tied undisputed world chess championship matches in chess history. Notice that most recent matches, Anand-Gelfand (2012), Karjakin-Carlsen (2016), and Carlsen-Caruana (2018) were not only very close, as seen in the columns labeled $\Delta$ and \%,  but the overall TPLVs were quite low compared with the earlier matches. Take the Schlechter-Lasker (1910) match, for example. It is worth noting that while Schlechter led the match 5.0-4.0 before the last round, he took risks in the last game to avoid a draw. However, this strategy did not pay off, and he lost the game. Therefore, the match ended in a tie, which meant that Lasker retained his title. (The reader may wonder how TPLVs apply in decisive matches. In the 1972 Fischer-Spassky ``Match of the Century,'' for example, Fischer significantly outperforms Spassky with a cumulative TPLV of 28.96 in decisive games compared to Spassky's 84.56. However, the cumulative TPLVs in the drawn games are similar, being 41.37 for Fischer and 41.29 for Spassky.)

\subsection{US Women's Championship}

Many elite tournaments, including the world chess championship as mentioned earlier, use Armageddon as a final tiebreaker. Most recently, the Armageddon tiebreaker was used in the 2022 US Women Chess Championship when Jennifer Yu and Irina Krush both tied for the first place, scoring each 9 points out of 13. Both players made big blunders in the Armageddon game; Irina Krush made an illegal move under time pressure and eventually lost the game and the championship. 

\begin{table}[ht!]
\centering
\[
\begin{array}{ r|c|c|c|}
\multicolumn{1}{r}{}
&  \multicolumn{1}{c}{\text{Player}}
& \multicolumn{1}{c}{\text{Cumulative TPLV}}
& \multicolumn{1}{c}{\text{Average TPLV}} \\
\cline{2-4}
&  \text{GM Irina Krush}  & 159.21 & 12.24  \\
\cline{2-4}
&  \text{GM Jennifer Yu}  & 188.62 & 14.50 \\
\cline{2-4}
\end{array}
\]
\caption{TPLV vs Armageddon tiebreakers in the 2022 US Women Chess Championship. Irina Krush would have been the champion because she had significantly lower cumulative TPLV in the tournament.}
\label{table:Krush-Yu-TPLV}
\end{table}

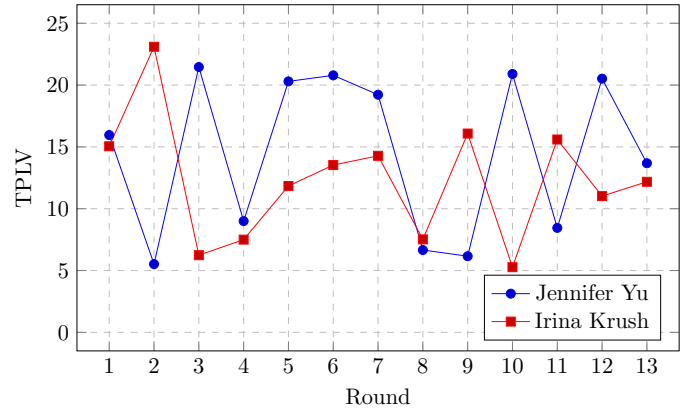
\begin{figure}
    \centering
\begin{tikzpicture}[scale=0.85]
\begin{axis}
[width=11cm, height=7cm,
    xlabel={Round}, 
    ylabel={TPLV}, 
    enlargelimits=0.06,
    xmin=1, xmax=13,
    ymin=0, ymax=25,
    xtick={1,2,3,4,5,6,7,8,9,10,11,12,13},
    ytick={0,5,10,15,20,25},
    grid,
    legend pos= south east,
    grid style=dashed]
\addplot 
	coordinates {(1, 15.96)
(2,5.52)
(3,21.46)
(4,9)
(5,20.3)
(6,20.79)
(7,19.22)
(8,6.66)
(9,6.16)
(10,20.9)
(11,8.45)
(12,20.52)
(13,13.68)};
\addplot 
	coordinates {(1, 15.05)
(2,23.1)
(3,6.24)
(4,7.5)
(5,11.84)
(6,13.53)
(7,14.28)
(8,7.52)
(9,16.08)
(10,5.27)
(11,15.6)
(12,11.02)
(13,12.18)};
\addlegendentry{Jennifer Yu}
\addlegendentry{Irina Krush}
\end{axis}
\end{tikzpicture}
    \caption{TPLVs of Irina Krush and Jennifer Yu in the 2022 US Women Chess Championship games. Lower TPLV implies better play.}
    \label{fig:Krush-Yu}
\end{figure}

Table~\ref{table:Krush-Yu-TPLV} presents the cumulative TPLVs of Irina Krush and Jennifer Yu in the 2022 US Women Chess Championship. More specifically, Fig~\ref{fig:Krush-Yu} illustrates their TPLVs for each round/game. According to our TPLV-based tiebreak method,  Irina Krush would have been the US champion because she played a significantly better chess in the tournament according to Stockfish: Irina Krush's games were about two pawn-units better on average than Jennifer Yu's.

\subsection{World Chess Championship 2018: Carlsen vs. Caruana}
\label{sec:Carlsen-Caruana}

Magnus Carlsen offered a draw in a better position against Fabiano Caruana in the last classical game in their world championship match in 2018. This was because Carlsen was a much better player in rapid/blitz time-control than his opponent. Indeed, he won the rapid tiebreaks convincingly with a score of 3-0. Note that Carlsen made the best decision given the championship match, but due to the tiebreak system his decision was not the best (i.e., manipulation-proof) in the particular game. This is related to the notion of strategyproofness that is often used in social choice, voting systems, mechanism design, and sports and competitive games; though, formal definition of strategyproofness varies depending on the context. For a selective literature, see, e.g., \cite{elkind2005,faliszewski2010,li2017,kendall2017,vong2017,aziz2018,brams2018c,dagaev2018,csato2019,pauly2013,chater2021,guyon2022,csato2022quantifying,csato2023b}.

Carlsen, of course, knew what he was doing when he offered a draw. During the post-game interview, he said ``My approach was not to unbalance the position at that point'' \cite{carlsen2018}. Indeed, in our opinion, Carlsen would not have offered a draw in their last game under the TPLV-based tiebreaking system because his TPLV was already lower in this game than Caruana's.

Carlsen's TPLV was 5.2 in the position before his draw offer, whereas Caruana's was 5.9. When Carlsen offered a draw, the evaluation of the position was about $-1.0$---i.e., Black is a pawn-unit better---according to Sesse, which is a strong computer running Stockfish. If Carlsen played the best move, then his evaluation would be about $-1.0$, which means he would be ahead about a pawn-unit. After the draw offer was accepted, the game ended in a draw and the evaluation of the position is obviously 0. As a result, Carlsen lost $$1~=~-(-1-0)$$ pawn-unit with his offer, as calculated in Table~\ref{table:Carlsen}. Thus, his final TPLV is $6.2=5.2+1$ as Table~\ref{table:Carlsen} shows. (For comparison, Carlsen's and Caruana's average centipawn losses in the entire match were 4.13 and 4.24, respectively.) A draw offer in that position would make Caruana the winner of the tiebreak under our method. As previously mentioned, we use TPLV to break a tie in matches; though, it could also be used as a tiebreaker in individual drawn games.

\begin{table}[ht!]
\centering
\[
\arraycolsep=1.1pt\def\arraystretch{1.5}
\begin{array}{ r|c|c|}
\multicolumn{1}{r}{}
&  \multicolumn{1}{c}{}
& \multicolumn{1}{c}{\text{GM Magnus Carlsen (B)}}\\
\cline{2-3}
&  \text{TPLV before draw offer}  & 5.2   \\
\cline{2-3}
&  \text{Evaluation of best move}  & -1.0  \\
\cline{2-3}
&  \text{Evaluation of draw offer}  & 0.0  \\
\cline{2-3}
&  \text{Pawn loss of draw offer}  & 1~=~-(-1-0)  \\
\cline{2-3}
&  \text{TPLV after draw offer}  & 6.2~=~5.2+1  \\
\cline{2-3}
\end{array}
\]
\caption{Calculation of TPLV after Carlsen's draw offer to Caruana. Negative values imply that the chess engine deems Carlsen's position better as he has the black pieces.}
\label{table:Carlsen}
\end{table}

\section{The definition, data collection, and computation}
\label{sec:chess}

\subsection{Definition}
\begin{table}
\[
\arraycolsep=1.3pt\def\arraystretch{2}
\begin{array}{ r|c|c|c|}
\multicolumn{1}{r}{}
&  \multicolumn{1}{c}{\text{Game Theory}}
& \multicolumn{1}{c}{\text{Notation}}
& \multicolumn{1}{c}{\text{Chess}}\\
\cline{2-4}
&  \text{a game}  & G & \text{the game of chess}  \\
\cline{2-4}
&  \text{a player}  &i\in \{1,2\} & \text{White or Black}  \\
\cline{2-4}
&  \text{an action}  & a_i\in A_i & ~\text{a move} ~ \\
\cline{2-4}
&  \text{a play}  & \bar{a}\in \bar{A} & \text{a single chess game} \\
\cline{2-4}
&  \text{a node}  & x_j\in X & \text{a position}  \\
\cline{2-4}
&  \text{a tournament}  & T(G) & \text{a tournament}  \\
\cline{2-4}
&  \text{AI}  & v_i: X\rightarrow \mathbb{R} & \text{a chess engine}  \\
\cline{2-4}
& ~  \text{AI best-response}  ~ & ~ a^*_i(x)\in \arg\max_{a_i(x)\in A_i(x)} v_i(a_i(x)) ~ & \text{best move}  \\
\cline{2-4}
\end{array}
\]
\caption{The terminology used in game theory and chess}
\label{table:terminology}
\end{table}

We focus on chess for the sake of clarity. For a formal definition of extensive-form games, see e.g. \cite{osborne1994}. Let $G$ denote the extensive-form game of chess under the standard  International Chess Federation (FIDE) rules. Table~\ref{table:terminology} summarizes the relationship between the terminologies used in game theory and chess. In the chess terminology, a \textit{chess game} is an alternating sequence of actions taken by White and Black from the beginning to the end of the game. In game theory, we call a chess game a \textit{play} and denote it by $\bar{a}\in \bar{A}$, which is the finite set of all plays. (For a recent paper that emphasizes focusing on plays rather than strategies in game theory, see \cite{bonanno2022}.) A chess \textit{position} describes in detail the history of a chess game up to a certain point in the game. Formally, in game theory, a position is a \textit{node}, denoted by $x\in X$,  in the extensive-form game $G$. A chess \textit{move} is a choice of a player in a given position. A \textit{draw offer} is a proposal of a player that leads to a draw if agreed by the opponent. If the opponent does not accept the offer, then the game continues as usual. Formally, an \textit{action} of a player $i\in \{1,2\}$ at a node $x\in X$ is denoted by $a_i(x)\in A_i(x)$, which is the set of all available actions of player $i$ at node $x$. An action can be a move, a draw offer, or the acceptance or rejection of the offer. A \textit{chess tournament}, denoted by $T(G)$, specifies the rules of a chess competition, e.g., the world chess championship where two players play a series of head-to-head chess games, and Swiss-system tournament where a player plays against a subset of other competitors.

We define a chess \textit{AI} as a profile $v$ of functions where for each player $i$, $v_i: X\rightarrow \mathbb{R}$. An AI yields an evaluation for every player and every position in a chess game. A chess engine is an AI which inputs a position and outputs the evaluation of the position for each player.

Let  $a^*_i\in A_i$ be an action at a node $x$. The action is called an \textit{AI best-response} if $a^*_i(x)\in \arg\max_{a_i(x)\in A_i(x)} v_i(a_i(x))$. In words, an AI best-response action at a position is the best move according to the chess engine $v$.

We are now ready to define ``pawn loss'' and total pawn loss.

\begin{definition}[Pawn loss]
Let $v_i(a^*_i(x_j))$ be a chess engine's evaluation of the best move for player $i$ at position $x_j$ and $v_i(a^j_i(x_j))$ be chess engine's evaluation of $i$'s actual move. Then, the \textit{pawn loss} of move $a^j_i(x_j)$ is defined as $v_i(a^*_i(x_j))-v_i(a^j_i(x_j))$.
\end{definition}

\begin{definition}[Total pawn loss value]

Let $\bar{a}\in \bar{A}$ be a chess game (i.e., a play) and $a^j_i$ be player $i$'s action at position $x_j$ in chess game $\bar{a}$, where  $\bar{a}_i=(a^1_i,a^2_i,...,a^{l_i}_i)$ for some $l_i$. Then, player $i$'s \textit{total pawn loss value} (TPLV) is defined as
\[TPLV_i(\bar{a}\ | \ v_i)=\sum_{j=1}^{l_i} [v_i(a^*_i(x_j))-v_i(a^j_i(x_j))].
\]

Let $\bar{a}^1, \bar{a}^2,..., \bar{a}^K$ where $\bar{a}^k\in \bar{A}$ be a sequence of chess games in each of which $i$ is a player. Player $i$'s \textit{cumulative} TPLV is defined as 
\[
C\text{-}TPLV_i((\bar{a}^1)^{}_{j=1})=\sum_{k=1}^{K} TPLV_i(\bar{a}^k\, |\,v_i).
\]
\end{definition}

In words, at every position the difference between the evaluations of a player's actual move and the best move is calculated. A player's TPLV is simply the total number of pawn-units the player loses during a chess game. TPLV is simply a mechanism that inputs an AI, $v$, and a chess game, $\bar{a}$, and outputs a tiebreak score for player $i$. 

It is straightforward to confirm that the average centipawn loss of a chess game $\bar{a}$ is given by $\frac{100}{l_i}\times TPLV_i(\bar{a}~ | ~ v_i)$. For both metrics, a zero value indicates that every move made by the player matches with the engine's top move. As mentioned in the introduction, we prefer using TPLV to average centipawn loss as a tiebreaker beucause (i) we find it more intuitive, and more importantly average centipawn loss can be manipulated by deliberately extending the game in  a theoretically drawn position to decrease one's own average centipawn loss.

The player with the lower $TPLV$ receives the higher tiebreaking score in a chess game.  This definition is extended to tiebreaking in chess tournaments (instead of tiebreaking in individual games) as follows. In the event of a tie in a chess tournament, the player(s) with the lowest  C-TPLV receive the highest tiebreaking score, and the players with the second lowest C-TPLV receive the second highest tiebreaking score, and so on.

\subsection{Further practicable tiebreaking rules}
We next define a specific and practicable AI tiebreaking rule. Let $\bar{a}$ be a chess game, $s_i$ the score of player $i$, and $TPLV_i<TPLV_j$ be player $i$'s TPLV in $\bar{a}$. If player $i$ wins the chess game $\bar{a}$, then $i$ receives 3 points and player $j\neq i$ receives 0 points: $s_i=3$ and $s_j=0$. If the chess game $\bar{a}$ is drawn, then each player receives 1 point. If $TPLV_i$ $<TPLV_j$, then the player $i$ receives an \textit{additional} 1 point, and player $j$ does not receive any additional point: $s_i=2$ and $s_j=1$. If $TPLV_i=TPLV_j$, then each player receives an additional $0.5$ points.

In simple words, we propose that the winner of a chess game receives 3 points (if they have a lower TPLV) and the loser 0 points, and in the event of a draw, the player with the lower TPLV receives 2 points and the other receives 1 point. This $(3,2,1)$ scoring system, which is currently used in many European ice hockey leagues, is akin to the scoring system used in volleyball if the match proceeds to the tiebreak, which is a shorter fifth set. Norway Chess also experimented with the $(3,2,1)$ scoring system, but now uses $(3,1.5,1)$ system perhaps to further incentivize winning a game. To our knowledge, Norway Chess was the first to use Armageddon to break ties at the game level rather than at the tournament level. 

There are several ways one could use TPLV to break ties. Our definition~ provides a specific tiebreaking rule in case of a tie in a chess game. For example, AI tiebreaking rule can also be used with the $(3,1.5,1)$ scoring system: The winner of a game receives 3 points regardless of the TPLVs, the winner of the tiebreak receives 1.5 points and the loser of the tiebreaker receives 1 point. In short, based on the needs and specific aims of tournaments, the organizers could use different TPLV-based scoring systems. Regardless of which tiebreaking rule is used to break ties in specific games, our definition provides a tiebreaking rule based on cumulative TPLV in chess tournaments. In the unlikely event that cumulative TPLVs of two players are equal in a chess tournament, the average centipawn loss of the players could be used as a second tiebreaker; if these are also equal, then there is a strong indication that the tie should not be broken. (For uniformity against slight inaccuracies in chess engine evaluations, we suggest using a certain threshold, within which the TPLVs can be considered equivalent.) But if the tie has to be broken in a tournament such as the world championship, then we suggest that players play two games---one as White and one as Black---until the tie is broken by the AI tiebreaking rule. In the extremely unlikely event that the tie is not broken after a series of two-game matches, one could, e.g., argue that the reigning world champion should keep their title.

\subsection{Data collection and computation}

As previously highlighted, our dataset encompasses all undisputed world chess championship matches in history that concluded in a tie (\url{www.pgnmentor.com}). Each match consists of a fixed, even number of head-to-head games, with players alternating colors. This format naturally leads to the possibility of ties. Our analysis reveals a notable trend: while only 11\% of matches ended in a tie before the year 2000, this figure rose to approximately 38\% in the post-2000 era, as illustrated in Fig~\ref{fig:wch}.

The significance of the world championship cannot be overstated; it represents the pinnacle of competitive chess, and winning the world championship is considered the ultimate achievement in a chess player's career. Our dataset is extensive, covering 286 games across nine world championship matches, with players making just above 25,000 moves in total.

Our primary goal is to determine a winner in tied matches by favoring the player with a lower cumulative TPLV, rather than resorting to playing speed chess tiebreaks. It is important to clarify that our objective is not to retroactively predict the actual winner of these historical matches. Instead, our approach provides a novel way to assess these games through systematic analysis.

For our analysis, we employed Stockfish 16, one of the most powerful chess AI systems available. This analysis was conducted on a personal desktop computer with the engine set to a depth of 20, signifying a search depth of about 20 moves (see \url{https://www.chessprogramming.org/Depth}). Note that, at these settings, Stockfish 16  significantly surpasses the strength of the best human players, providing generally a reliable and objective assessment of each move. For this reason, our settings are reasonable in comparison with those reported in the literature; for example, \cite{backus2023} uses Houdini 1.5a at a depth of 15, and \cite{kunn2022} uses Stockfish 11 at a depth of 25.

\begin{figure}[h]
    \centering
 \includegraphics[width=0.8\textwidth]{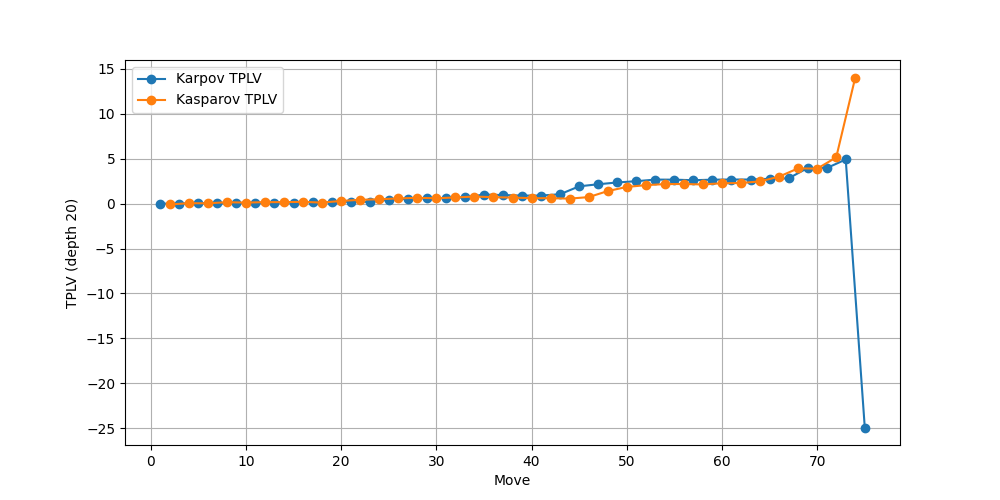}
\includegraphics[width=0.8\textwidth]{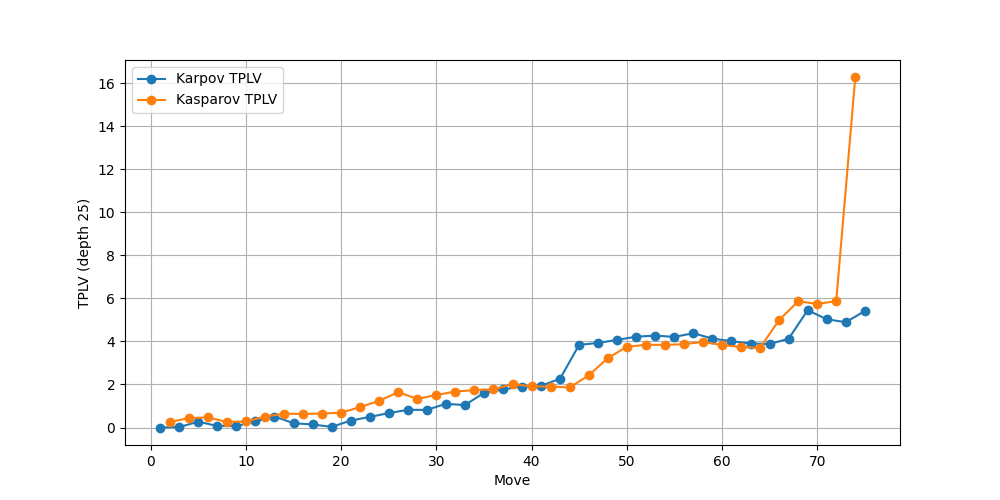}
\caption{TPLV dynamics in game 5 of the 1987 Kasparov-Karpov match under different depths}
\label{fig:game5}
\end{figure}

An exceptional instance occurred in game 5 of the Kasparov-Karpov match, where Karpov's TPLV was approximately -200. This suggested he played at a level surpassing the Stockfish depth 20 setting, a rare but not unheard-of occurrence. Consequently, we reanalyzed the entire match at a depth of 25 (see Fig~\ref{fig:game5}). We have pinpointed the move that caused this anomaly. Under depth 20, Karpov's last move significantly increased his evaluation of the position, an unexpected outcome as the best engine move should not change the evaluation, and a suboptimal move should decrease it. This indicates that, at this setting, the engine was unable to evaluate the position correctly one move earlier. However, when analyzed at depth 25, this anomaly disappeared, and the outcome of the tiebreak, according to our method, remained unchanged.

We have included all game-level data in the Supporting Information, specifically in Fig~\ref{fig:all}, and Figs~\ref{fig:outcomes1} and \ref{fig:outcomes2}. Table~\ref{table:summary} presents summary statistics. The table includes the years the matches took place, the players, the total number of games in each match, and the cumulative TPLV (C-TPLV) for each player.

\section{Limitations}
\label{sec:limitations}

\subsection{Sensitivity to the software (AI system) and the hardware} 

Both the AI system (software) and the hardware play a role in calculating TPLVs in a game. In addition, as noted earlier, specific software settings, such as search depth, can also affect the evaluations produced by the selected AI system. Consequently, it is essential that all factors influencing the evaluations be disclosed publicly prior to a tournament.

In particular, the engine settings should be kept fixed across all games, unless the tournament director has a reasonable doubt that the AI's assessment of a particular position in a game was flawed in a way that might affect the result. In that case, the tournament director may seek a re-evaluation of the position/game. 

Today, several of the best chess engines, including Stockfish and AlphaZero, are widely acknowledged to be clearly much better than humans. Thus, either of these chess engines could be employed for the AI tiebreaking rule. (In tournaments with a large number of participants, however, one could use computationally less expensive engine settings to calculate the TPLVs.)

\subsection{Sensitivity to the aggregation of TPLVs and the threshold selection}

\begin{figure}[h]
    \centering
    \begin{subfigure}[b]{0.5\textwidth}
        \centering
        \includegraphics[width=\textwidth]{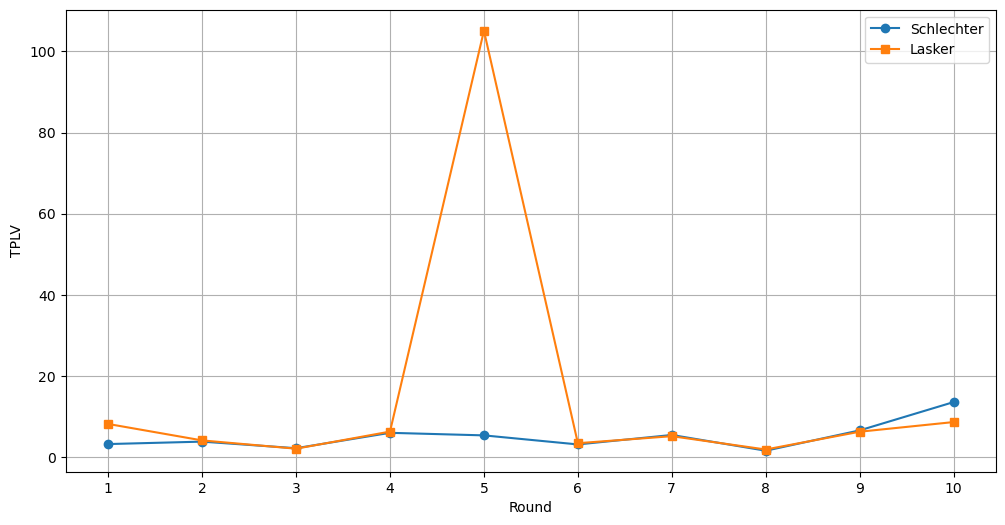}
        \caption{Schlechter-Lasker 1910}
        \label{fig:lasker}
    \end{subfigure}%
    \begin{subfigure}[b]{0.5\textwidth}
        \centering
        \includegraphics[width=\textwidth]{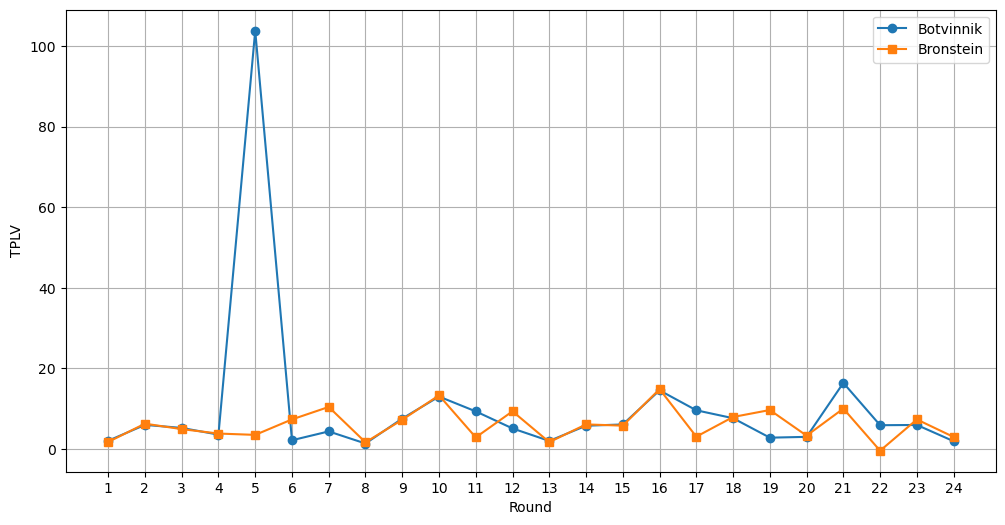}
        \caption{Botvinnik-Bronstein 1951}
        \label{fig:botvinnik}
    \end{subfigure}
    \begin{subfigure}[b]{0.5\textwidth}
        \centering
        \includegraphics[width=\textwidth]{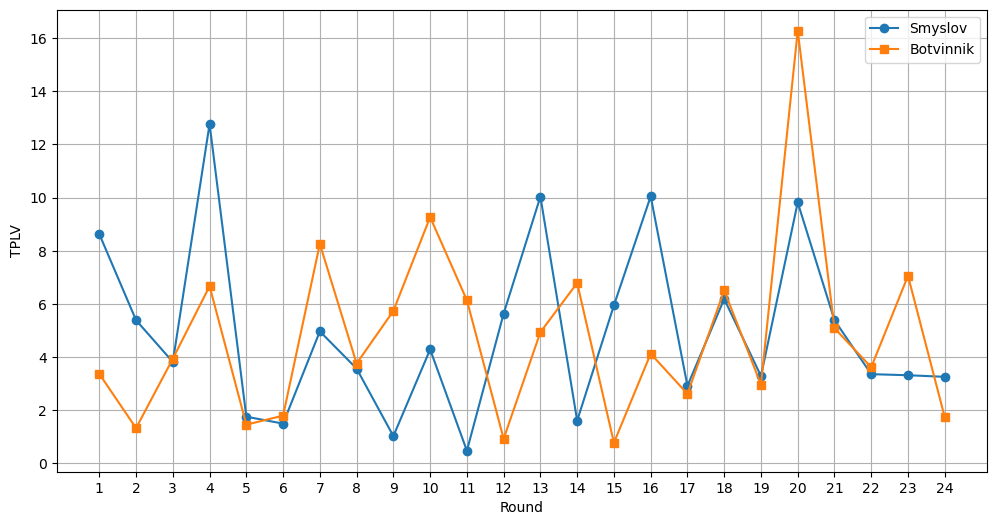}
        \caption{Botvinnik-Smyslov 1954}
        \label{fig:smyslov}
    \end{subfigure}%
    \begin{subfigure}[b]{0.5\textwidth}
        \centering
        \includegraphics[width=\textwidth]{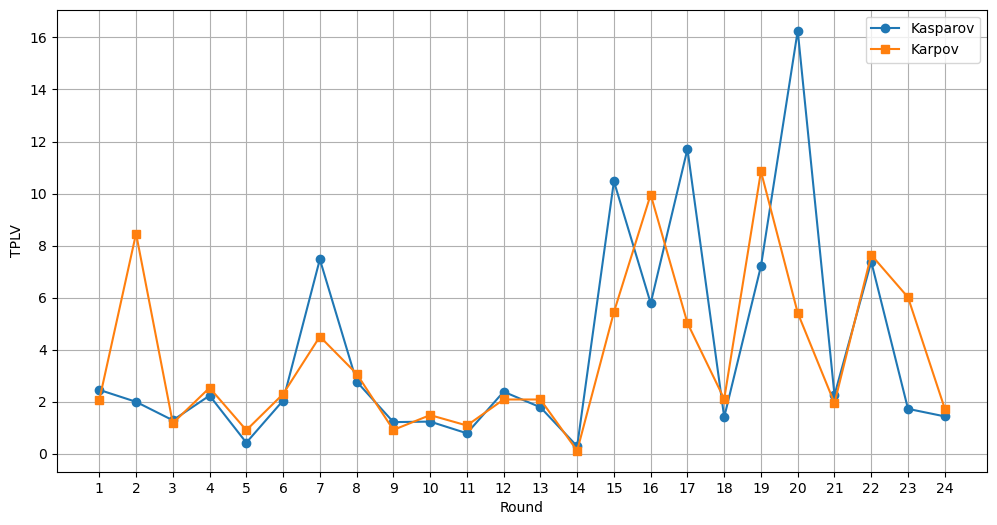}
        \caption{Kasparov-Karpov 1987}
        \label{fig:karpov}
    \end{subfigure}
    \begin{subfigure}[b]{0.5\textwidth}
        \centering
        \includegraphics[width=\textwidth]{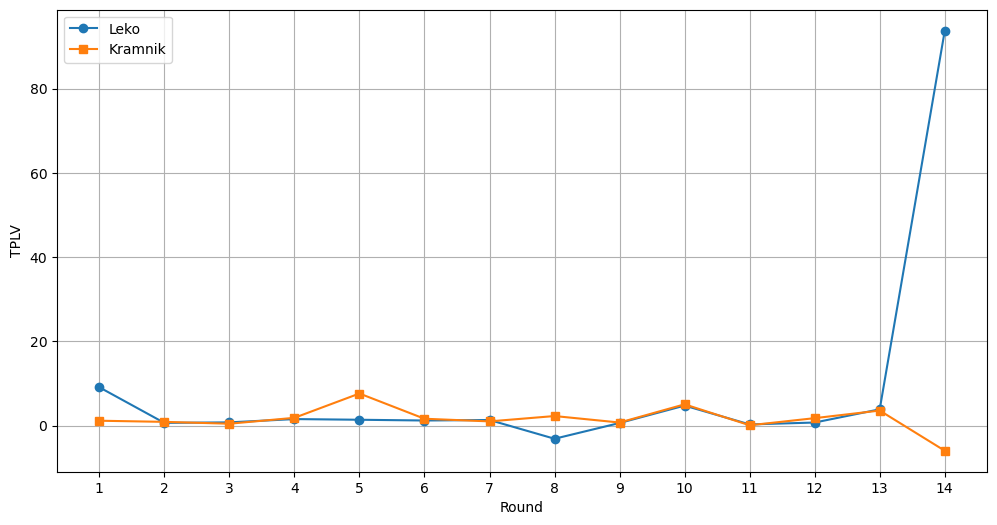}
        \caption{Leko-Kramnik 2004}
        \label{fig:leko}
    \end{subfigure}%
    \begin{subfigure}[b]{0.5\textwidth}
        \centering
        \includegraphics[width=\textwidth]{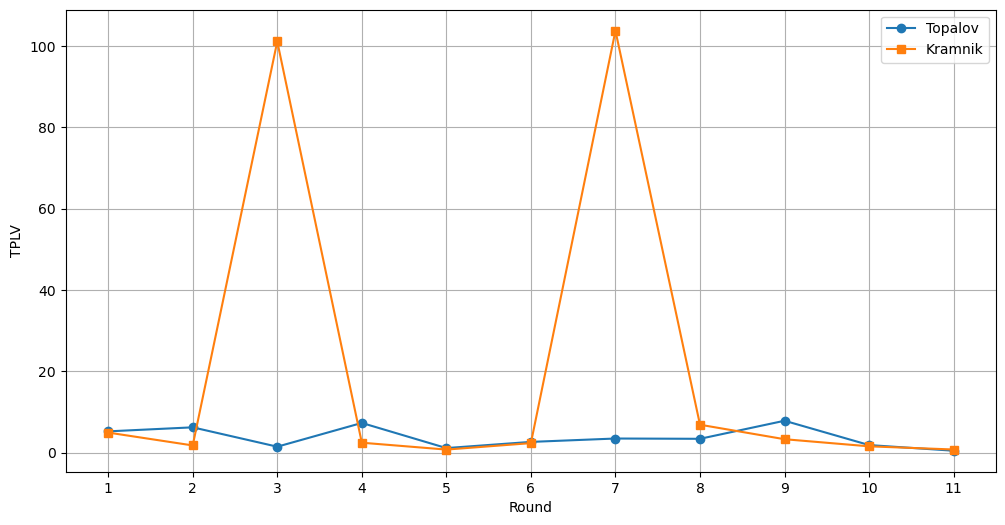}
        \caption{Topalov-Kramnik 2016}
        \label{fig:topalov}
    \end{subfigure}
    \begin{subfigure}[b]{0.5\textwidth}
        \centering
        \includegraphics[width=\textwidth]{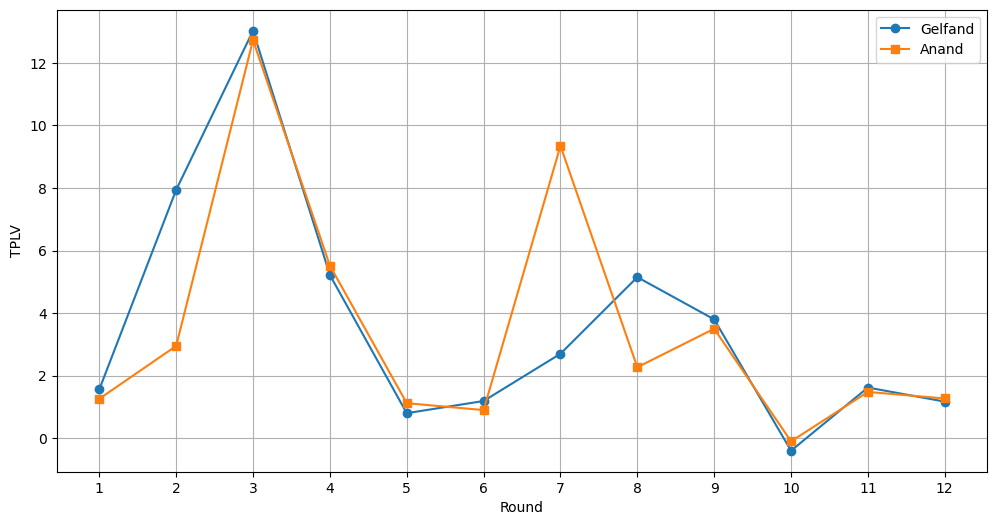}
        \caption{Anand-Gelfand 2012}
        \label{fig:anand}
    \end{subfigure}%
    \begin{subfigure}[b]{0.5\textwidth}
        \centering
        \includegraphics[width=\textwidth]{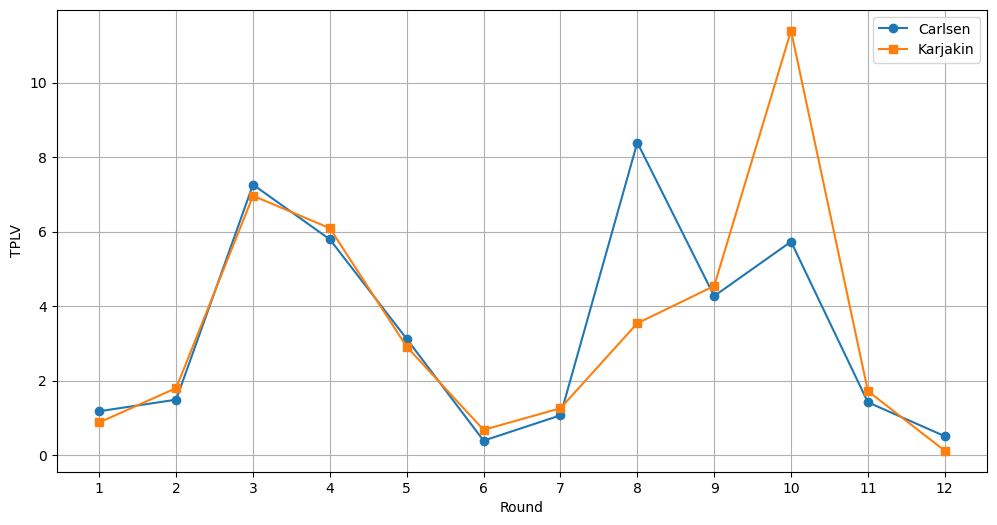}
        \caption{Karjakin-Carlsen 2016}
        \label{fig:carlsen}
    \end{subfigure}
    \begin{subfigure}[b]{0.5\textwidth}
        \centering
        \includegraphics[width=\textwidth]{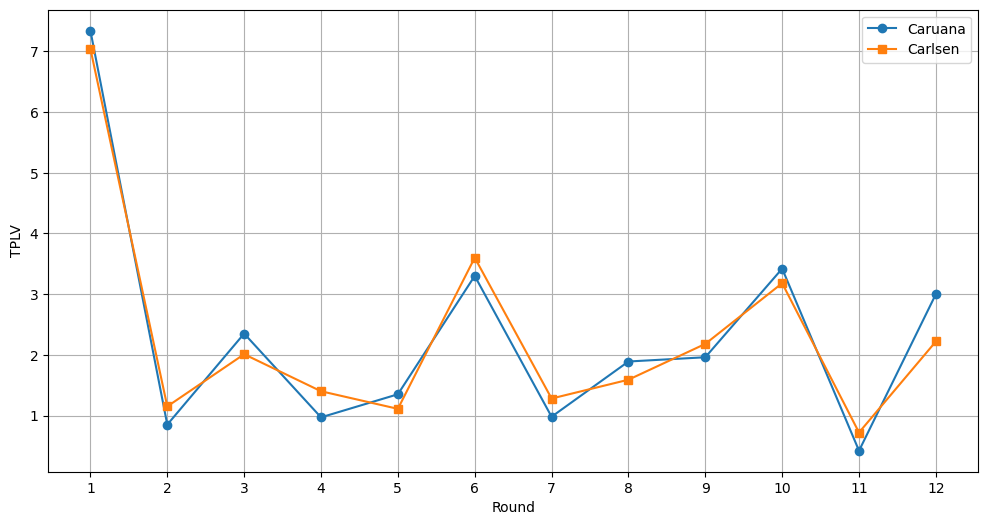}
        \caption{Carlsen-Caruana 2018}
        \label{fig:carlsen2}
    \end{subfigure}
\caption{TPLVs per round/game in World Chess Championship matches}
\label{fig:all}
\end{figure}

As indicated in Table~\ref{table:summary}, the player with the lowest cumulative TPLV wins the tiebreak in a match. A plausible objection to this tiebreak mechanism is that aggregating evaluations at every position in each game may disproportionately benefit one player over another. For example, consider the Schlechter-Lasker (1910) and Botvinnik-Bronstein (1951) matches shown in Fig~\ref{fig:all}. In each of these matches, there is only one game in which the TPLVs of the players are significantly different, whereas in all other games, the TPLVs are close. Should a single game, or even a single move, determine the outcome of the tiebreaker? 

Different sports competitions have experimented with alternative approaches to aggregation issues. For example, in Rubik's cube competitions, one's worst result and best result are excluded before taking the average completion time. A similar approach could be adopted in chess: each player's worst and best games (and/or moves) could be excluded before calculating the cumulative TPLVs. Naturally, such a change should be announced prior to the event.

Furthermore, an argument can be made that  the player with the higher TPLV should win the tiebreak in the event of a tie, rather than the player with the lower TPLV. This argument stems from the observation that the player with the lower TPLV failed to capitalize on the mistakes of the opponent. However, this approach conflicts with the following method for breaking ties. Consider a tournament where player A and player B draw their game; player A wins all other games with perfect play, whereas player B, despite winning the remaining games, makes many blunders. In this case, it is arguably more natural to break the tie in favor of player A.

Finally, we have suggested that a threshold might be set, whereby two TPLVs that fall within this fixed threshold are considered equivalent. Alternatively, one could set a percentage threshold whereby if the difference between TPLVs is within this threshold with respect to, say, the lower TPLV, then the TPLVs are considered equivalent. For example, with a threshold of 1\%, TPLVs of 100 and 101 would be considered equivalent. To illustrate, with a threshold percentage of 1\%, all ties can be broken in Table~\ref{table:summary}. However, with a threshold percentage of 5\%, the ties in the Botvinnik-Smyslov, Anand-Gelfand, Karjakin-Carlsen, and Carlsen-Caruana matches would remain unbroken.

\subsection{Different playing styles}
The playing style---positional vs tactical, or conservative vs aggressive---of a player may make them more (or less) susceptible to making mistakes against a player with a different playing style. A valid concern is whether our AI mechanisms favor one style over another. The answer depends on the chess engine (software) and the hardware that are used to break ties. A top player may have a better ``tactical awareness'' than a relatively weak chess engine or a strong engine that runs on a weak hardware. Using such an engine to break ties would then obviously be unfair to the player.
However, there is little doubt that the latest version of Stockfish running on a strong hardware is a better tactical and/or positional player than a human player. As an analogy, suppose that a world chess champion evaluates a move in a game played by amateur players. While the world champion may be biased, like any other player, there is little doubt that their evaluation would be more reliable than the evaluation of an amateur.

Another point of concern could be the impact of openings that typically result in ``theoretical draws,'' such as the Berlin draw. There is an argument to be made for excluding such games from the TPLV calculations. Exploring how TPLV-based tiebreaking can be adapted in such cases is an interesting direction for future research. Another factor that could potentially change the playing style of the players occurs when they are informed of the TPLVs during a game. If this were possible, it would greatly impact players because giving top players evaluations of positions would help them considerably, even without the TPLV considerations. However, this would violate the fair play rules.

\subsection{Computer-like play}
A reasonable concern could be that our proposal will make players to play more ``computer-like.'' Nevertheless, top chess players now play arguably more like engines than they did in the past, as indicated by the apparent decrease in cumulative TPLVs shown by Table~\ref{table:summary}. Expert players try to learn as much as they can from engines, including openings and end-game strategies, in order to gain a competitive edge. As an example, \cite{carlsen2022} 
recently explained how he gained a huge amount of knowledge and benefited from neural network-based engines such as AlphaZero. He also said that some players have not used these AIs in the correct way, and hence have not benefited from them. (For a further discussion, see \cite{gonzalez2022}.) In summary, there is little, if anything, that the players can do to play more computer-like and take advantage of the AI tiebreaking mechanism on top of what they normally do. To put it slightly differently, if there is any ``computer-like'' chess concept that a player can learn and improve their AI score, then they would learn this concept to gain a competitive edge anyway---even if AI tiebreaking mechanism is not used to break ties. That being said, it is up to the tournament organizers to decide which chess engine to use for tiebreaking, and some engines are more ``human-like'' than the others \cite{mcilroy2020}.

\subsection{Playing strength}

It is simpler to play (and win) a game against a weaker opponent than a stronger opponent, and a player is less likely to make mistakes when playing against a weaker opponent. Is it then unfair to compare the quality of the moves of different players? We do not think so. First, in most strong tournaments, including the world championship and the candidates tournament, every player plays against everyone else. Second, in Swiss-system tournaments, players who face each other at any round are in general of comparable strength due to the format of this tournament. While it is impossible to guarantee that each tied player plays against the same opponents in a Swiss-system tournament, we believe that AI tiebreaking mechanisms are arguably preferable to other mechanisms because they are impartial and based on the quality of the moves played by the player himself or herself. (For ranking systems used in Swiss-system tournaments and a review of these systems, see, e.g., \cite{csato2013,csato2017}.)

\subsection{How will the incentives of the players change?}

Apart from boosting the quality of matches by naturally giving more incentives to players to find the best moves, our quality-based tiebreaking rule provides two additional benefits. First, observe that it is very likely to discourage ``prematurely'' agreed draws, as there is no assurance that each player will have the same TPLV when a draw is agreed upon during a game; thus, at least the player who senses having the worse (i.e., higher) TPLV up to that point will be less likely to offer or agree to a draw. Second, this new mechanism is also likely to reduce the incentive for players to play quick moves to ``flag'' their opponent's clock---so that the opponent loses on time---because in case of a draw by insufficient material, for instance, the player with the lower TPLV would gain an extra point.

\subsection{What is the ``best strategy'' under our tiebreak mechanism?}

Another valid question is whether playing solid moves, e.g., the top engine moves from beginning to the end, is the ``best strategy'' under any of our tiebreak mechanisms. The answer is that the best strategy in a human vs human competition is \textit{not} to always pick the top engine moves. This is because the opponent might memorize the line that is best response to the top engine moves in which case the outcome of the game would most likely be a draw. Under our tiebreak mechanism, the only time a player should deviate from the optimal move should be when one does \textit{not} want the game to go to the tiebreak---i.e., when one wants to win the game. And, winning the game is more valuable than winning a tiebreak. Therefore, playing a sub-optimal computer move might be the ``human-optimal'' move to win the game.

\section{Conclusions}
\label{sec:conclusions}

In contrast to the current tiebreak system of rapid, blitz, and Armageddon games, the winner of the tiebreak under a quality-based tiebreak strategyproof AI mechanism is determined by an objective, state-of-the-art chess engine with an Elo rating of about 3600. Under the new mechanism, players' TPLVs are highly likely to be different in the event of a draw by mutual agreement, draw by insufficient material, or any other `regular' draw. Thus, for a small enough threshold value, nearly every game will result in a winner, making games more exciting to watch and thereby increasing fan viewership.

A valid question for future research direction is whether and to what extent our proposal could be applied to other games and sports. While our tiebreaking mechanisms are applicable to all zero-sum games, including chess, Go, poker, backgammon, football (soccer), and tennis, one must be cautious when using such an AI tiebreaking mechanism in a game/sport where AI's superiority is not commonly recognised, particularly by the best players in that game. Only after it is established that AI is capable of judging the quality of the game---which is currently the case only in a handful of games including Go, backgammon, and poker \cite{brown2019}---do we recommend using our tiebreaking mechanisms.

\section*{Supporting information}
S1 Data.

\begin{figure}[htbp]
    \centering
\resizebox{\textwidth}{!}{%
\begin{tikzpicture}

\newcommand{\cross}{\tikz[scale=0.2]{
    \draw[thick] (0,0) -- (1,1);
    \draw[thick] (0,1) -- (1,0);
}}

\foreach \x in {1,...,14} {
    \node at (\x,1) {\x};
}

\node at (-1,0) {Schlechter-Lasker};
\foreach \x/\y in {1/0.5, 2/0.5, 3/0.5, 4/0.5, 5/1, 6/0.5, 7/0.5, 8/0.5, 9/0.5, 10/0} {
    \ifdim \y pt = 1 pt
        \node[draw,circle,inner sep=2pt,fill=black] at (\x,0) {};
    \else
        \ifdim \y pt = 0 pt
            \node at (\x,0) {\cross};
        \else
            \node[draw,circle,inner sep=2pt,fill=white] at (\x,0) {};
        \fi
    \fi
}

\node at (-1,-2) {Leko-Kramnik};
\foreach \x/\y in {1/0, 2/0.5, 3/0.5, 4/0.5, 5/1, 6/0.5, 7/0.5, 8/1, 9/0.5, 10/0.5, 11/0.5, 12/0.5, 13/0.5, 14/0} {
    \ifdim \y pt = 1 pt
        \node[draw,circle,inner sep=2pt,fill=black] at (\x,-2) {};
    \else
        \ifdim \y pt = 0 pt
            \node at (\x,-2) {\cross};
        \else
            \node[draw,circle,inner sep=2pt,fill=white] at (\x,-2) {};
        \fi
    \fi
}

\node at (-1,-4) {Topalov-Kramnik};
\foreach \x/\y in {1/0, 2/0, 3/0.5, 4/0.5, 5/1, 6/0.5, 7/0.5, 8/1, 9/1, 10/0, 11/0.5, 12/0.5} {
    \ifdim \y pt = 1 pt
        \node[draw,circle,inner sep=2pt,fill=black] at (\x,-4) {};
    \else
        \ifdim \y pt = 0 pt
            \node at (\x,-4) {\cross};
        \else
            \node[draw,circle,inner sep=2pt,fill=white] at (\x,-4) {};
        \fi
    \fi
}

\node at (-1,-6) {Anand-Gelfand};
\foreach \x/\y in {1/0.5, 2/0.5, 3/0.5, 4/0.5, 5/0.5, 6/0.5, 7/0, 8/1, 9/0.5, 10/0.5, 11/0.5, 12/0.5} {
    \ifdim \y pt = 1 pt
        \node[draw,circle,inner sep=2pt,fill=black] at (\x,-6) {};
    \else
        \ifdim \y pt = 0 pt
            \node at (\x,-6) {\cross};
        \else
            \node[draw,circle,inner sep=2pt,fill=white] at (\x,-6) {};
        \fi
    \fi
}

\node at (-1,-8) {Karjakin-Carlsen};
\foreach \x/\y in {1/0.5, 2/0.5, 3/0.5, 4/0.5, 5/0.5, 6/0.5, 7/0.5, 8/1, 9/0.5, 10/0, 11/0.5, 12/0.5} {
    \ifdim \y pt = 1 pt
        \node[draw,circle,inner sep=2pt,fill=black] at (\x,-8) {};
    \else
        \ifdim \y pt = 0 pt
            \node at (\x,-8) {\cross};
        \else
            \node[draw,circle,inner sep=2pt,fill=white] at (\x,-8) {};
        \fi
    \fi
}

\node at (-1,-10) {Carlsen-Caruana};
\foreach \x/\y in {1/0.5, 2/0.5, 3/0.5, 4/0.5, 5/0.5, 6/0.5, 7/0.5, 8/0.5, 9/0.5, 10/0.5, 11/0.5, 12/0.5} {
    \ifdim \y pt = 1 pt
        \node[draw,circle,inner sep=2pt,fill=black] at (\x,-10) {};
    \else
        \ifdim \y pt = 0 pt
            \node at (\x,-10) {\cross};
        \else
            \node[draw,circle,inner sep=2pt,fill=white] at (\x,-10) {};
        \fi
    \fi
}

\end{tikzpicture}    
}
\vspace{0.1cm}
\caption{Outcomes of world championship matches with 14 or fewer classical games: a hollow dot represents a draw, a solid dot represents a win for the first player, and a cross represents a loss for the first player}
    \label{fig:outcomes1}
\end{figure}

\begin{figure}
    \centering
\resizebox{\textwidth}{!}{%
\begin{tikzpicture}

\newcommand{\cross}{\tikz[scale=0.2]{
    \draw[thick] (0,0) -- (1,1);
    \draw[thick] (0,1) -- (1,0);
}}

\node at  (-1,0) {Botvinnik-Bronstein};
\foreach \x/\y in {1/0.5, 2/0.5, 3/0.5, 4/0.5, 5/0, 6/1, 7/1, 8/0.5, 9/0.5, 10/0.5, 11/0, 12/1, 13/0.5, 14/0.5, 15/0.5, 16/0.5, 17/0, 18/0.5, 19/1, 20/0.5, 21/0, 22/0, 23/1, 24/0.5} {
    \ifdim \y pt = 1 pt
        \node[draw,circle,inner sep=2pt,fill=black] at (\x,-2) {};
    \else
        \ifdim \y pt = 0 pt
            \node at (\x,-2) {\cross};
        \else
            \node[draw,circle,inner sep=2pt,fill=white] at (\x,-2) {};
        \fi
    \fi
}

\node at  (-1,-2) {Botvinnik-Smyslov};
\foreach \x/\y in {1/1, 2/1, 3/0.5, 4/1, 5/0.5, 6/0.5, 7/0, 8/0.5, 9/0, 10/0, 11/0, 12/1, 13/1, 14/0, 15/1, 16/1, 17/0.5, 18/0.5, 19/0.5, 20/0, 21/0.5, 22/0.5, 23/0, 24/0.5} {
    \ifdim \y pt = 1 pt
        \node[draw,circle,inner sep=2pt,fill=black] at (\x,0) {};
    \else
        \ifdim \y pt = 0 pt
            \node at (\x,0) {\cross};
        \else
            \node[draw,circle,inner sep=2pt,fill=white] at (\x,0) {};
        \fi
    \fi
}

\node at (-1,-4) {Kasparov-Karpov};
\foreach \x/\y in {1/0.5, 2/0, 3/0.5, 4/1, 5/0, 6/0.5, 7/0.5, 8/1, 9/0.5, 10/0.5, 11/1, 12/0.5, 13/0.5, 14/0.5, 15/0.5, 16/0, 17/0.5, 18/0.5, 19/0.5, 20/0.5, 21/0.5, 22/0.5, 23/0, 24/1} {
    \ifdim \y pt = 1 pt
        \node[draw,circle,inner sep=2pt,fill=black] at (\x,-4) {};
    \else
        \ifdim \y pt = 0 pt
            \node at (\x,-4) {\cross};
        \else
            \node[draw,circle,inner sep=2pt,fill=white] at (\x,-4) {};
        \fi
    \fi
}

\foreach \x in {1,...,24} {
    \node at (\x,1) {\x};
}
\end{tikzpicture}

}
\vspace{0.1cm}
\caption{Outcomes of world championship matches with 24 classical games: a hollow dot represents a draw, a solid dot represents a win for the first player, and a cross represents a loss for the first player}
    \label{fig:outcomes2}
\end{figure}

\section*{Acknowledgments}
We would like to thank FIDE World Chess Champion (2005-2006) and former world no. 1 Grandmaster Veselin Topalov for his comments and encouragement: ``I truly find your idea quite interesting and I don't immediately find a weak spot in it. There might be some situations when the winner of the tiebreak does not really deserve it, but that's also the case with any game of chess'' (e-mail communication, 11.08.2022,  from Topalov regarding our tiebreaking proposal). We are also grateful to Steven Brams, Michael Naef and Shahanah Schmid for their valuable comments and suggestions.


%
%
%


\end{document}